\documentclass[12pt,preprint]{aastex}

\shorttitle{Lifetime of clumps in the ISM}
\shortauthors{Falceta-Gon\c calves \& Lazarian}

\begin{document}

\title{Evolution and lifetime of transient clumps in the turbulent interstellar medium}
\author{D. Falceta-Gon\c calves\altaffilmark{1,2} \& A. Lazarian\altaffilmark{2}}
\altaffiltext{1}{Escola de Artes, Ci\^encias e Humanidades, Universidade de S\~ao Paulo - 
Rua Arlindo Bettio 1000, CEP: 03828-000, S\~ao Paulo, Brazil }
\altaffiltext{2}{Astronomy Department, University of Wisconsin, Madison, 475 N. Charter 
St., WI 53711, USA}

\begin{abstract}
We study the evolution of dense clumps and provide argument that the existence of the clumps is not limited by
the crossing time of the clump. We claim that the lifetimes of the clumps are determined 
by the turbulent motions on larger scale and predict the correlation of the clump 
lifetime and its column density. 
We use numerical simulations and successfully test this relation. In addition, we study 
the morphological asymmetry and the magnetization of the clumps as a function of their 
masses.  
\end{abstract}
\keywords{ISM: clouds, kinematics and dynamics - stars: formation - methods: numerical -
MHD}

\section{Introduction}

Star formation is believed to take place in collapsing cores within molecular clouds. 
These dense structures however are turbulent and magnetized. Both physical properties, 
allied to the gas thermal pressure, help in the support of the cores against 
gravitational collapse. During the past years many have worked on this interesting 
subject and many issues remained. One important but still open issue regards the true 
lifetimes of dense clumps in the interstellar medium (ISM). Evaporation, instabilities 
and the turbulence itself are believed to destroy and fragment the clumps at short 
timescales, comparable to the sonic/Alfvenic crossing timescales. On the other hand, 
gravitationally unstable cores should also fragment and collapse at even shorter 
timescales $< 10^5$yrs. These short timescales, on the other hand, contradicts 
observations of star forming regions and also turns the formation of massive stars 
virtually impossible. 

The interstellar medium (ISM) is known to be turbulent (Armstrong et al. 1995, Elmegreen 
\& Scalo 2002, Chepurnov \& Lazarian 2010) with interstellar gas exhibiting a wide variety of structures, including 
giant molecular clouds at large scales, and cloudlets or clumps at small scales. Star 
formation is believed to be triggered within these dense clumps and, therefore, the 
understanding of the formation of these structures and their dynamical evolution is 
crucial for theories of star formation and its efficiency in the ISM.

In the fragmentation model (Elmegreen \& Mathieu 1983) it is assumed that 
the molecular clouds collapse due to their gravity, as the thermal energy is 
continuously reduced by the emission of radiation. During this process the gas 
temperature decreases while the density increases, and the Jeans mass, $m_{\rm 
J}(M_{\odot}) \simeq 10 n^{-1/2} 
T^{3/2}$, being the temperature $T$ in Kelvin and the gas density $n$ in $cm^{-3}$, is 
reduced resulting in cloud fragmentation. This process should occur 
up to scales where the opacity becomes too large and the radiative cooling 
inefficient. In such a scenario the 
dynamics of the plasma plays little role. However, it 
is well known that molecular clouds are formed in supersonic turbulence, and the typical 
physical energies may be estimated as:

\begin{eqnarray}
E_{\rm grav} \simeq \frac{3Gm_c^2}{5R} \sim 10^{47}\ {\rm erg}, \nonumber \\
E_{\rm thermal} \simeq \frac{2\pi nkTR^3}{3} \sim 10^{47}\ {\rm erg}, \nonumber \\
E_{\rm turb} \simeq \frac{1}{2} m_c c_s^2 M_s^2 \sim 3 \times 10^{46} M_s^2\ {\rm erg},
\end{eqnarray}

\noindent
for a typical giant molecular cloud of $m_c = 10^4$M$_{\odot}$, $T=20$K, $R=25$pc, where 
$c_s$ is the isothermal speed of sound and $M_s = u_{\rm turb}/c_s$ 
the averaged sonic Mach number of the turbulent eddies. For molecular clouds, 
observations provide typical values of $M_s \simeq 3 - 20$ 
(Larson 1981, Li \& Houde 2008), resulting in supersonic turbulence at most of molecular 
clouds. Though gravity continues to be the main cause 
of clump contraction and star formation, the distribution of cloud sizes and masses 
may depend mostly on the turbulence itself. From the theoretical point of view, 
magneto-hydrodynamical (MHD) 
simulation have been extensively used for studying the statistics of density distribution 
in turbulent plasmas (e.g. Ballesteros-Paredes, V\'azquez-Semadeni \& Scalo 1999; Padoan, 
Goodman \& Juvela 2003; Kowal, Lazarian \& Beresniak 2007; Burkhart et al. 2009), all 
presenting 
density and column density PDFs similar to the observed, i.e. with density contrasts 
between the large scale cloud and the dense cores of $\rho_{\rm core}/<\rho> \sim 10^2 - 
10^4$, in the supersonic turbulent flows.  

These dense structures that are not gravitionally dominated are believed to be 
transient, and most important, short lived (Kirk, Ward-Thompson \& Andre 2005). 
A fraction of these clumps are believed to present masses larger than the Jean's one 
being subject to fast collapse, being the collapsing free-fall time estimated as $t_{\rm 
ff} = 3.4 \times 10^7 n^{-0.5} \sim 10^4-10^5$yrs. This value is too short compared to 
the lifetimes required for the formation of high mass stars ($t > 10^5$yrs) (Churchwell 
1999, Mac Low \& Klessen 2004). For this reason, several supporting mechanisms, other 
than thermal pressure, have been proposed. Observationally, the turbulent flows in 
molecular clouds are supersonic at scales $l > 0.05$pc and have been 
invoked as the main support mechanisms against the gravitational collapse of the clouds 
(Bonazzola et al. 1987, Xie et al. 1996). Magnetic fields also play a major role 
in this process, while the uniform component of the field supports the cloud in the 
perpendicular direction (Mestel \& Spitzer 1956, McKee \& Zweibel 1995; Gammie \& Ostriker
1996; Martin, Heyvaerts \& Priest 1997), the perturbations
(Alfvenic mode mostly) being responsible 
for the extra pressure in the parallel direction (see Falceta-Gon\c calves, de Juli \& 
Jatenco-Pereira 2003). In both cases, since turbulence also excites the Alfvenic modes 
within molecular clouds, turbulence must exist for $t > 10^5$yrs. One important issue is 
that turbulence decays in a relatively short timescale $t_{\rm damp} \sim l/v \sim 10^3 - 
10^4$yrs, where $l\sim 0.1$pc is the turbulent cell size and $v\sim 5$km s$^{-1}$ its 
turnover speed. Clouds that are not collapsing are believed to be quickly destroyed as 
well, either by 
instabilities such as Kelvin-Helmholtz (Kamaya 1996), shocks and evaporation. This fact 
motivates a study of the dynamics of clump evolution. What is the origin of these dense 
regions? How long do the clumps survive in a turbulent medium? 

In this work we study the formation, and the dynamical evolution of the transient dense 
regions in molecular clouds based on numerical simulations. We concentrate this paper on 
the survival of clumps to shocks and turbulent diffusion. We also discuss, based on the 
simulations, the origin and morphology of the identified dense structures. In Section 2 
we describe the code and basic assumptions used in order to obtain the results, shown in 
Section 3. In Section 4 we discuss the main results and present the conclusions of this 
work, followed by the Summary.

\section{Code description}

The gas in starless molecular clouds may be considered isothermal, though selfgravity may 
play a major role in its dynamics. In this work however we will disregard the 
gravitational force and consider exclusively the effects of turbulence in the formation 
and destruction of dense clumps.

The simulations were performed solving the set of ideal MHD isothermal equations, in 
conservative form, as follows:

\begin{equation}
\frac{\partial \rho}{\partial t} + \mathbf{\nabla} \cdot (\rho{\bf v}) = 0,
\end{equation}

\begin{equation}
\frac{\partial \rho {\bf v}}{\partial t} + \mathbf{\nabla} \cdot \left[ \rho{\bf v v} + 
\left( p+\frac{B^2}{8 \pi} \right) {\bf I} - \frac{1}{4 \pi}{\bf B B} \right] = {\bf f},
\end{equation}

\begin{equation}
\frac{\partial \mathbf{B}}{\partial t} - \mathbf{\nabla \times (v \times B)} = 0,
\end{equation}

\begin{equation}
\mathbf{\nabla \cdot B} = 0,
\end{equation}

\begin{equation}
p = c_s^2 \rho,
\end{equation}

\noindent
where $\rho$, ${\bf v}$ and $p$ are the plasma density, velocity and pressure, 
respectively, ${\bf B = \nabla \times A}$ is the magnetic field, ${\bf A}$ is 
the vector potential and ${\bf f} = {\bf f_{\rm turb}}$ represents the external source 
terms, such as the turbulence forcing. The code solves the set of MHD equations 
using a Godunov-type scheme, based on a second-order-accurate and essentially 
non-oscillatory spatial reconstruction. We use the HLL Riemann 
solver for shock-capturing and the magnetic divergence is assured by the 
constrained transport method for the induction equation. The code has been extensively 
tested (Falceta-Gon\c 
calves, Lazarian \& Kowal 2008; Le{\~a}o et al. 2009; Burkhart et al. 2009; Kowal et al. 
2009; Falceta-Gon\c calves et al. 2010a,b).
The turbulence is triggered by the injection of solenoidal perturbations in Fourier space of the velocity field.
  
We run 6 models with numerical resolution of $512^3$ cells, in a fixed uniform grid, at 
different turbulent regimes, ranging from subsonic to supersonic and sub to 
super-Alfvenic, which were already presented in Falceta-Gon\c calves, Lazarian \& Houde 
(2010) in a different study. The box size corresponds to $L_{box} = 10$pc, the 
assumed average number density of the gas $n_{\rm H} = 10^2$cm$^{-3}$, the magnetic 
field strength $B \sim 2\mu$G and $T \sim 20$K.

\section{Results}

The numerical simulations show a very dynamical picture with clumps changing their 
location 
and shapes. As discussed in Kowal, Lazarian \& Beresnyak (2007), the density contrast 
$\rho / \bar{\rho}$ follows a proportionality with the sonic Mach number, with little  
dependence on the Alfvenic Mach number. As the goal of this work is on the 
dynamics of the dense structures we will focus our calculations on the supersonic case 
(Model \# 3 of Falceta-Gon\c calves, Lazarian \& Houde 2010). This specific run presented 
average sonic Mach number $M_{\rm s} = 7.2$, and showed $\rho / \bar{\rho} \sim 1000 - 
5000$ depending on the snapshot. 
The initial magnetic field is set uniform with $\beta = P_{\rm thermal}/P_{\rm mag} = 
1$.
The dynamical evolution of clumps was studied based on 100 snapshots per dynamical time 
$t_{\rm D} = L/v_{\rm L}$, and the simulation performed up to $t = 5 t_{\rm D}$.

\subsection{Identification and origin of clumps}

Basically, we identify the clumps in a very simple way, using isocontours of density. The 
clump identification method determines, for each snapshot, all cells of the simulated 
cube with density above a given threshold $\rho'_{\rm min} = \rho / \bar{\rho}$. All 
other cells, with $\rho 
< \rho'_{\rm min}$ are flagged and not used. The program then identifies as a single 
clump the region with all neighboring overdense cells. This procedure is similar to 
the one described in Schmidt et al. (2010), with the difference that the limit of the 
clump size is given by it total mass, rather than the boundary density. In Figure 
1, we show the clumps 
identified at $t=4 t_{\rm D}$ for two different thresholds $\rho'_{\rm min} = 20$ and 30.

\begin{figure}
\centering
\includegraphics[scale=0.2]{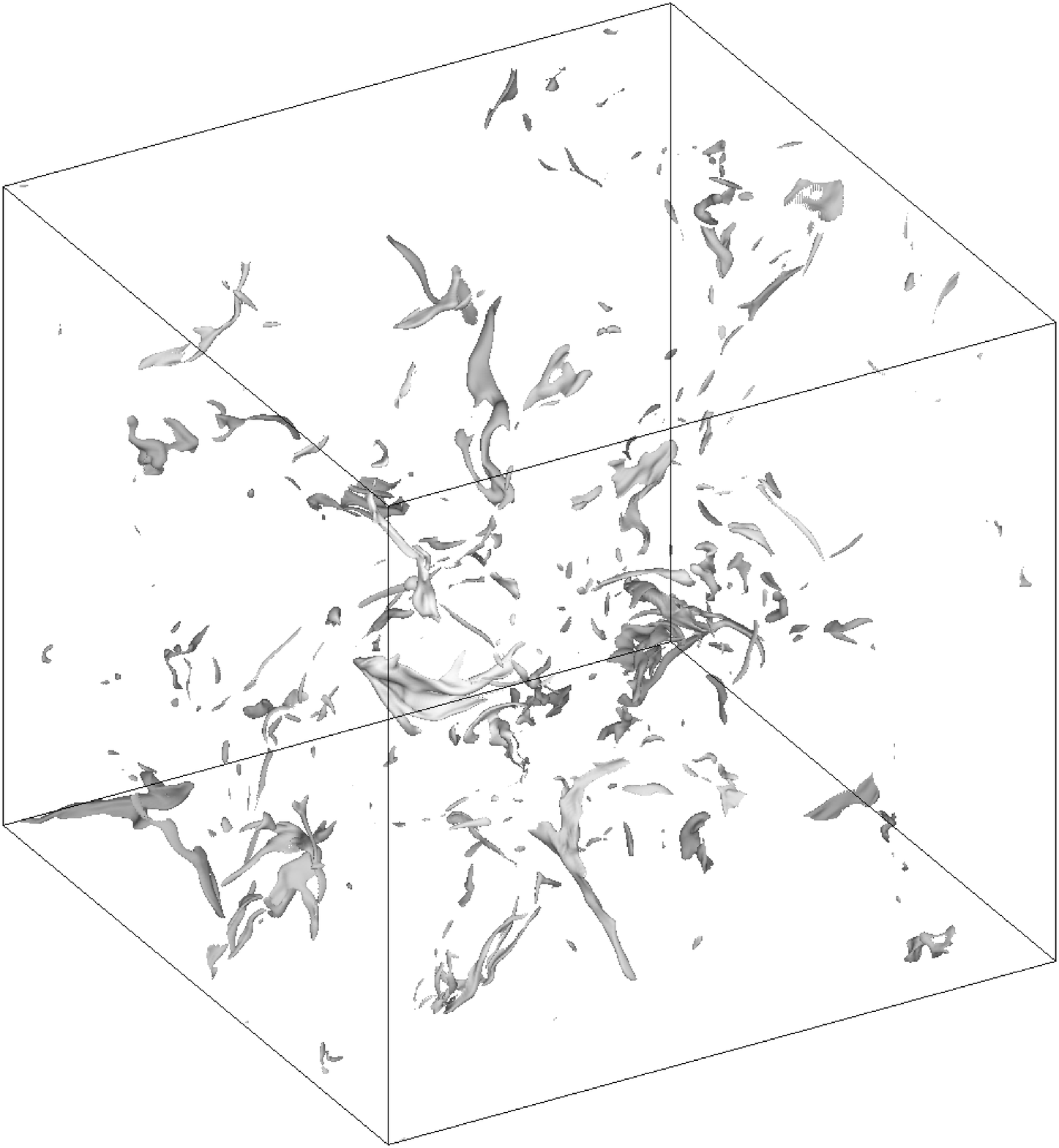} \\
\includegraphics[scale=0.2]{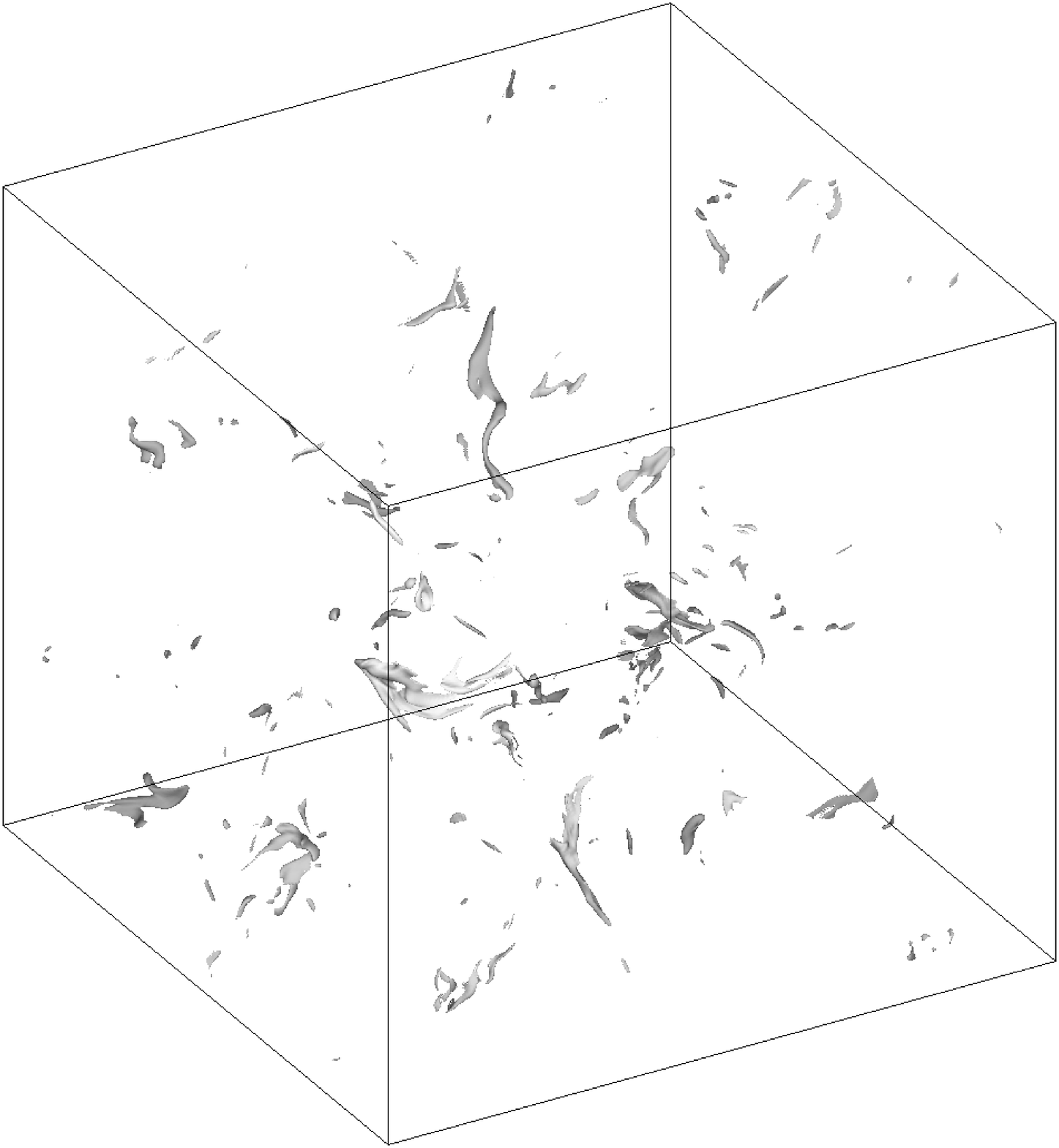}  
\caption{Visualization of the overdense regions of the simulation at snapshot $t=4 t_{\rm 
D}$ for thresholds $\rho'_{\rm min} = 20$ and 30.}
\label{clumps}
\end{figure}

At each snapshot different clumps, varying both in size and location, are identified. In 
order to follow the dynamics of each clump it is mandatory that we determine the 
displacement of a clump from $t_1=t'$ to the subsequent frame $t_2=t'+\Delta t$. 
Our method uses the local averaged velocity field to estimate future positions of the 
clump. At $t=t'$, for each clump, the average velocity ${\bf u} = \Sigma_N {\bf v_i} / N$ 
is obtained. The estimated location of each cell of the given clump is then obtained by 
${\bf x_i} = {\bf x_i^0}+{\bf u}\Delta t$. At snapshot $t_2$, the same procedure is 
executed. At $t_2$ if a clump - or part of it - is identified at within the region 
defined 
by ${\bf x_i}$ it is flagged as an old clump that is simply evolving. If more than one 
clump is identified in the region defined by ${\bf x_i}$, then two different 
possibilities are considered: i- the clump has fragmented, or ii- a different clump moved 
and occupied this region. If the second condensation is not located also at other set 
of ${\bf x_i}$, determined from a different clump at $t=t'$, it may be considered a 
fragment.

\begin{figure*}
\centering
\includegraphics[scale=0.8]{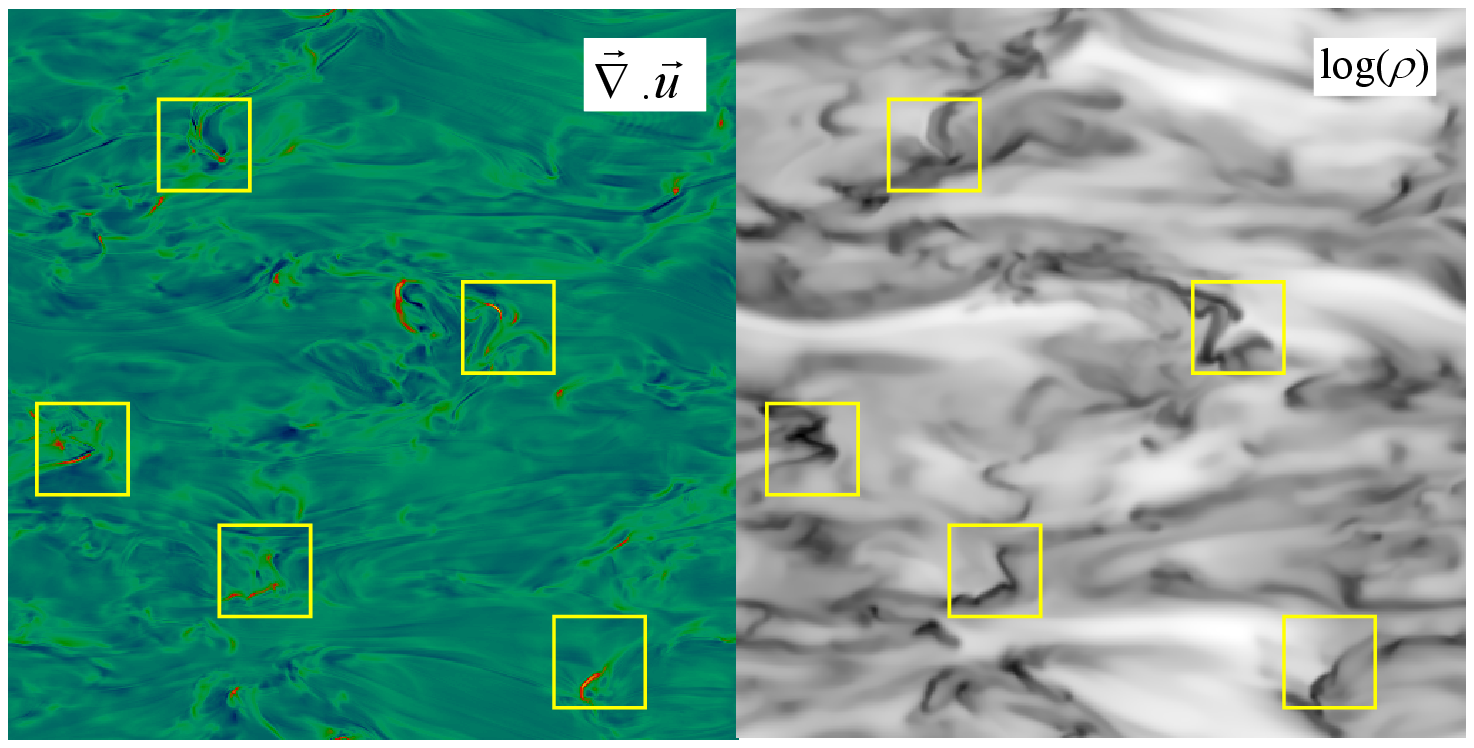}   
\caption{For a central slice at $t=4 t_{\rm D}$, {\it left panel}: Divergence of the 
velocity field, and {\it right panel} logarithm of the gas density. Boxes are used to 
show the regions where we observe obvious correspondence of structures. }
\label{shocks}
\end{figure*}

From the velocity field it is also possible to determine the origin of the 
condensations. 
Transient clumps may be produced by accumulating matter in a supersonic flow. 
The converging flows of the gas may have different origins, including locally ''colliding 
flows"\footnote{the proper idea is that of converging flows as ''colliding flows" may 
induce the reader to the erroneous impression that separate jet-like flows are commom in 
developed turbulence.} (Elmegreen 
1993, Blitz \& Williams 1999) or isothermal shocks (Beresnyak, Lazarian \& Cho 2005). 
In Figure 2, we show two maps. One is for divergence of velocity, and the 
other shows density peaks. The correlation between clarifies the origin of the gas 
clumpiness. The strong divergence of velocity at the peaked density (mostly at the edges) 
indicate that shocks are responsible for them. Visually, one can verify that the 
majority of the clumps are associated with shocks. At the center of the image, a vertical 
shock is formed but uncorrelated to a density peak. The reason for this is its 
evolutionary stage. In subsequent frames the density increases at the shock region is 
recognized.

\subsection{Morphology and magnetic fields}

Additionally, the clump finding routine was developed to measure the clump dimensions and 
morphology. Two dimensions are determined: the lengths along and perpendicular to the 
mean magnetic field within the condensation. The asymmetry of the cloud is calculated as 
the ratio between the two lengths $A=L_{\parallel}/L_{\perp}$. The average ratio of the 
gas and magnetic pressures within the cloud, $\beta$, is also obtained for each 
structure. The result is shown in Figure 3 for
all clumps found by this method. In the figure, M is the mass of the clump, normalized 
by solar masses.

\begin{figure}
\centering
\includegraphics[scale=0.65]{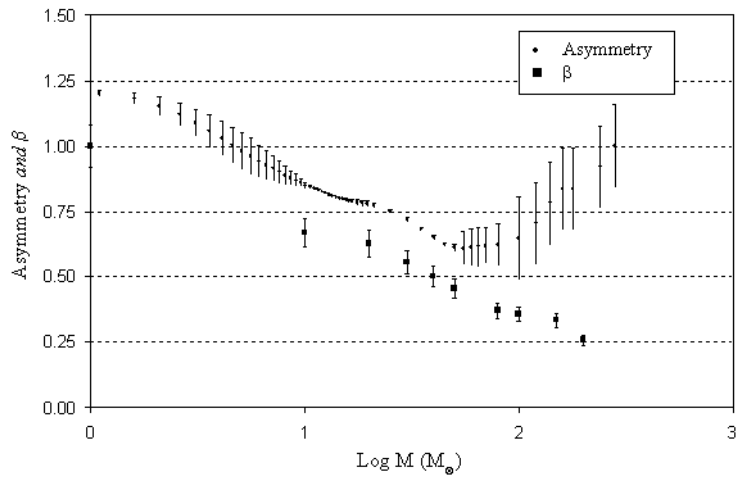}   
\caption{Asymmetry of dimensions parallel and perpendicular to the mean magnetic 
field ($A=L_{\parallel}/L_{\perp}$), and averaged $\beta = P_{\rm th}/P_{\rm mag}$, for 
the clumps with  respect to the total mass of the clump in solar masses.}
\label{simetry}
\end{figure}

We observe that most clumps are found to be compressed along the mean magnetic field, 
i.e. they are larger in the perpendicular direction. However, there is a selection effect 
with the cloud mass. Clumps with small mass ($M_{\rm clump} < 3$M$_{\odot}$) 
tend to be larger in the parallel direction and the most massive cores tend to be 
isotropic ($M_{\rm clump} > 200$M$_{\odot}$). Following the theory for compressible MHD turbulence the 
Alfvenic part of perturbations scales as  
$l_{\parallel} \sim l_{\perp}^{2/3}$ (see Goldreich \& Sridhar 1995, and numerical 
tests of Cho \& Lazarian 2002, 2003, Kowal \& Lazarian 2010). At the smaller scales the 
low velocity (subsonic), since $v_l \sim l^{1/3}$, 
results in a coupling between the density and velocity fields. Since the velocity fields 
are anisotropic and aligned with the local field, the density acts as an almost passive 
scalar aligning to the magnetic field lines as well. At intermediate/larger scales the 
yet supersonic motions result in the gas compression. The shock surfaces are mostly 
perpendicular to the magnetic field for two reasons: i- the shocks parallel to the field 
are weakened by the magnetic pressure, resulting in smaller contrast of density, and ii- 
the velocity anisotropy results in larger velocity components at parallel direction that 
produced stronger shocks, with density surfaces perpendicular to the field lines. At the 
other end, i.e. the most massive cores, the turbulence is strongly super-Alfvenic and the 
field mostly random. In this case, neither the velocity nor the the density present 
anisotropic distribution. 

Figure 3 also presents the averaged $\beta$ value 
for all clumps followed in this simulation. There is a clear anticorrelation between the 
total mass of the clump and $\beta$. An interesting effect that we observe is that the asymmetry of clumps and their 
$\beta$ for the range of masses 
go in a very similar fashion, i.e. they decrease with the increase of the mass of a 
clump, except for larger masses. We may interpret this as an 
effect of shearing of clumps by Alfvenic motions. We started the run with equal 
thermal and 
magnetic pressures, i.e. $\beta=1$, the magnetic pressure tend to increase within the 
clumps making them magnetically dominated. Turbulent 
motions, in general, result in compression of the magnetic field with $B \propto 
n^{\alpha}$, being $\alpha \sim 0.5 - 1.0$ (Burkhart et al. 2009). Since $P_{\rm th} 
\propto n$ and $P_{\rm mag} \propto B^2$ the more compressed the gas the lower is 
$\beta$.

\subsection{Clump lifetimes}

How long do the clumps live? The answer to this question is important for many 
astrophysical issues, such as core collapse, star formation and SF efficiency. 
As mentioned above, it has been frequent in the past years to invoke the shortest of the  
sonic and Alfvenic crossing times as an estimation of the cloud survival timescale,  
which is then compared to the free-fall or ambipolar diffusion timescales. 

\begin{figure}
\centering
\includegraphics[scale=0.55]{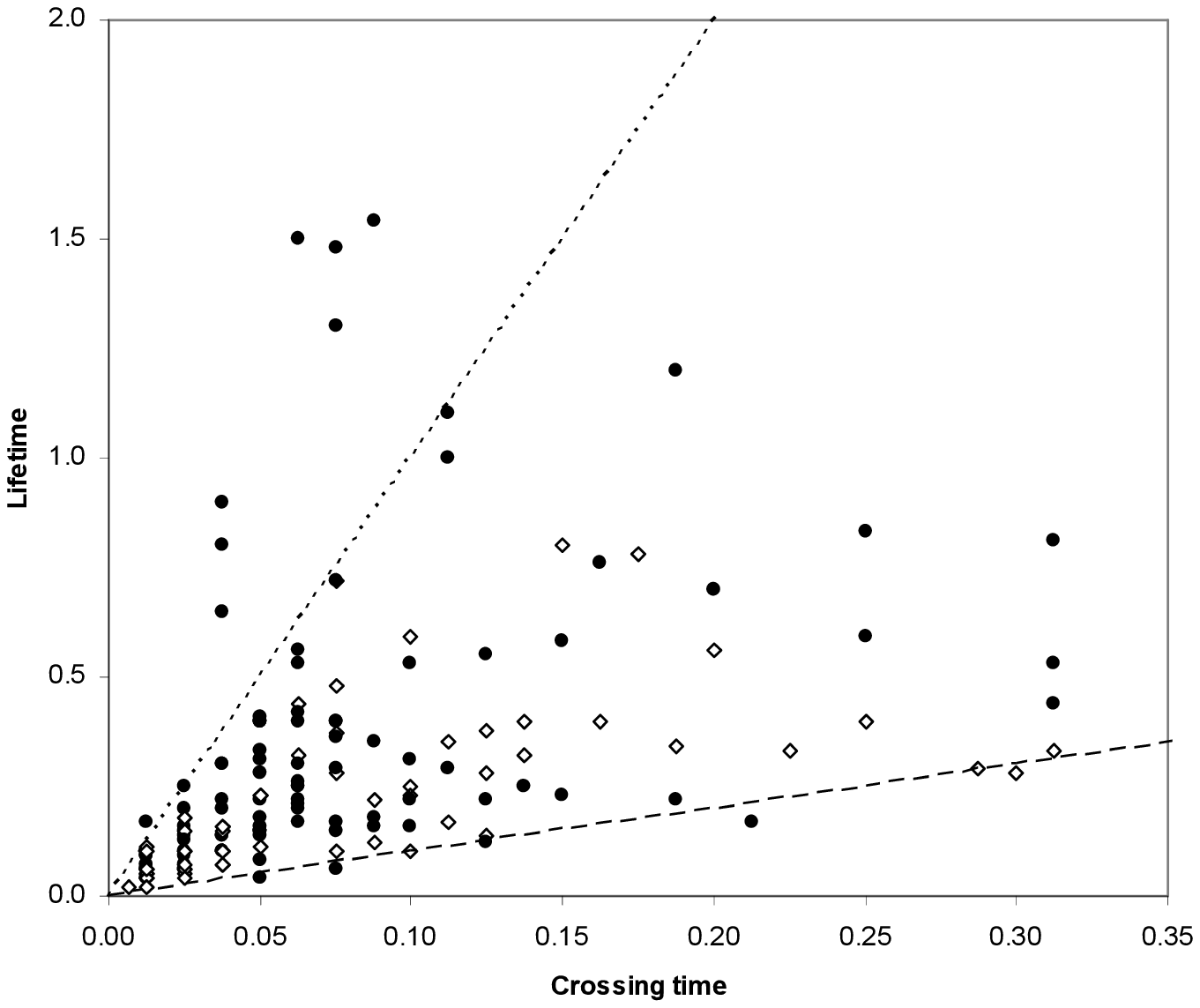}   
\caption{Survival times determined for all the cores identified for both thresholds 
$\rho'_{\rm min} = 20$ (open circles) and 30 (filled circles). The lifetimes are given in 
terms of the sonic crossing time $\tau_{\rm s} = l_{\rm clump} / c_{\rm s}$, which is 
shown in the abcisse. Dashed line 
corresponds to $t_{\rm life} = \tau_{\rm s}$ and dotted line corresponds to 
$t_{\rm life} = 10\tau_{\rm s}$.}
\label{timescales}
\end{figure}

From the simulation it was possible to track the evolution of individual clumps, 
determining the times of formation and destruction. In Figure 
4 we show the cores identified for both thresholds $\rho'_{\rm min} = 20$ 
(open circles) and 30 (filled circles). The dashed line corresponds to $t_{\rm 
life} = \tau_{\rm s}$, where $\tau_{\rm s} = l_{\rm clump} / c_{\rm s}$, and the dotted 
one corresponds to $t_{\rm life} = 10\tau_{\rm s}$. There is a correlation between the 
two timescales showing that 
larger cores are long-lived compared to the smaller ones. Also, the same conclusion is 
valid for the denser cores, as we compare two cores with similar sizes (similar $\tau_{\rm s}$) but with different density thresholds. The filled circles ($\rho'_{\rm min} = 
30$) populates an upper 
region in this plot compared to the open circles ($\rho'_{\rm min} = 20$).

As we could see in Figure 3, the clumps tend
to be magnetically dominated. Therefore, we expect the shortest timescale to be 
$t_A=l/V_A$, where $l$ is the clump size and $V_A$ is the Alfv\'en velocity. However, we 
see in Figure 4 that most of the clumps live longer than the 
sonic crossing time $\tau_{\rm s}$. In fact, most of the clumps live between 1 
and $10 \tau_s$, which, if we combine the results on plasma $\beta$ of clumps translates 
into the range from 1 to $20 t_A$.

This result is specially important for the formation of dense and massive cores. As 
mentioned above, for the ISM the collapse timescales of dense cores (even considering the 
ambipolar diffusion) tend to be fast enough to prevent the accumulation of material to 
form e.g. massive stars. It 
is important that it survives longer than $\sim t_{\rm cross}$.
A rough estimate of the gas accumulating in a high Mach number flow can be obtained by 
assuming that the compressible turbulent motions of size $l$ and 
density $\rho_l$ accumulate gas as

\begin{equation}
\rho v_L \tau_l \sim \rho_l l,
\label{accum}
\end{equation}

\noindent
where $v_L$ is the velocity of the flow at the energy injection scale, $\rho$ is the 
density of the unperturbed gas. 
This provides an estimate for the time for accumulating the material of the clump:

\begin{equation}
\tau_l \sim \frac{\rho_l l}{\rho v_L},
\label{time}
\end{equation}

\noindent
which indicates that the characteristic life time of the clump is actually determined by 
the longer time scales associated with the compressible flows. 

Examining Eq.(8) we see that, contrary to the wide-spread belief the clumps may 
survive much
longer than the sound crossing time. We see that clumps may accumulate matter for as long 
as the large
scale converging flow exist. The latter may proceed up to the injection time $L/v_L$. In the model proposed the 
accumulated column density of a clump is proportional to $\rho_l l$ and is also 
proportional to the time the compressible converging flow accumulates the material. 

To test this simple idea we calculate the column density of each of the clumps 
and relate it with their lifetimes. In order to obtain an estimation of the column 
density - that does not depend on any specific orientation of a given line of sight - we 
calculated the averaged dimension of each cloud and multiplied by its averaged density, 
i.e. $\Sigma \sim \bar{n}_{\rm c} \bar{l}_{\rm c}$. The result is shown in Figure 
5 in real physical units. The 
solid line shows a good linear relationship between the two quantities, $\tau_{\rm life} 
\propto \Sigma$. The dispersion is expected due to the crude nature of our definition of 
a clump. A formal fit to the data using linear regression gives $\alpha=0.72 \pm 
0.08$. The power index obtained from the statistical fit is similar to our prediction. 
If only the clumps with threshold $\rho'_{\rm 
min} = 20$ are considered the linear regression gives $\alpha=0.81 \pm 0.12$, even closer to our
prediction. The discrepancy between the predicted scaling and the statistical fit may be 
due to a completeness 
effect. Since it is possible that there is a lack of clumps in the top-right region of 
the plot, i.e. denser clumps that survive longer, simply because of limited numerical 
resolution that artificially introduce a large viscosity in the system. As a consequence, 
the small clumps are more easily destroyed. Further simulations with finer resolution, or 
better schemes, are recommended in order to test this possibility.

\begin{figure}
\centering
\includegraphics[scale=0.55]{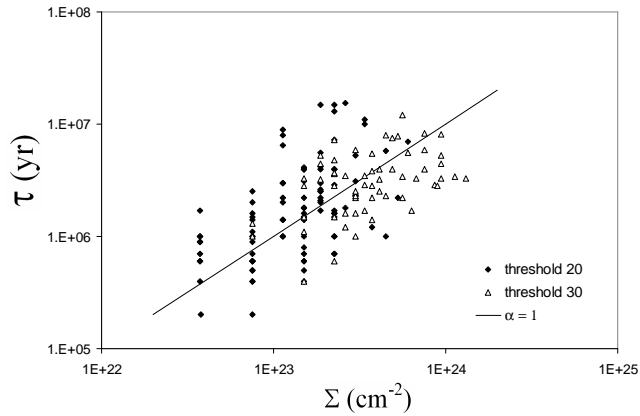}   
\caption{Relationship between the lifetime of a clump and its column density. The 
solid line is given as a reference and corresponds to a linear relationship 
$\tau_{\rm life} \propto \Sigma^a$, being a=1.}
\label{cdens}
\end{figure}

\subsection{Clump evaporation vs turbulent disruption}

The dense clumps formed within molecular clouds may be subject to gas evaporation. This 
process happens once there is no pressure equilibrium between the denser core and the 
surrounding material of the cloud. In the case of starless cores we may consider the 
molecular cloud plasma shielded to background radiation, and the gas approximately 
isothermal. The contrast of density is then responsible for a pressure gradient resulting 
in evaporation. In such case thermal conduction is not the dominant physical 
process for the gas evaporation but the gas diffusion, i.e. the flow from the cores to 
the less dense regions.

A crude estimate of the clump evaporation rate may be obtained (see Cowie \& McKee 1977). 
We will assume a spherical clump, in a quasi-stationary approximation, that follows the 
momentum equation given as:

\begin{equation}
\rho v \frac{\partial v}{\partial r} \simeq -\frac{\partial P}{\partial r},
\end{equation}

\noindent
where $\rho$ is the gas mass density, $P$ the pressure and $r$ is the distance to the 
center of the clump. Assuming the gas at the core and its surrounding as isothermal, as 
explained above, this equation is simplified as follows:

\begin{equation}
\frac{\partial \ln(\rho)}{\partial r} = -\frac{1}{2} \frac{\partial M^2}{\partial r},
\end{equation}

\noindent
where $M$ is the sonic Mach number of the gas flow, and gives the solution:

\begin{equation}
M^2 = 2 \ln \left( \frac{\rho_0}{\rho} \right).
\end{equation}

The evaporation timescale is given by:

\begin{equation}
\tau_{\rm evap} \simeq \frac{m_{\rm c}}{\dot{m}_{\rm evap}},
\end{equation} 

where $m_{\rm c}$ is the total mass of the clump and $\dot{m}_{\rm evap}$ the mass loss 
rate,

\begin{equation}
\dot{m}_{\rm evap}=4 \pi r^2 \rho v.
\end{equation}

Assuming a gaussian radial distribution for the gas density within the clump, the 
evaporation timescale is reduced to,

\begin{equation}
\tau_{\rm evap} \simeq \frac{\sqrt{2}}{3} \frac{r_{\rm c}^4}{r^3} \frac{1}{c_{\rm s}} 
\exp\left( \frac{r^2}{2r_{\rm c}^2} \right) \sim 3 \times 10^{6} {\rm yrs},
\end{equation}

\noindent
for a typical clump size obtained in the simulation of $r_{\rm c} = 0.3$pc, with $c_{\rm 
s} = 1$km s$^{-1}$, and $r/r_{\rm c} = 1/e$. This evaporation timescale of the clump is 
much larger compared to other dynamical scales of the systems, such as the sound  
and turbulent crossing times ($\sim 10 \tau_{\rm sound}$ and $\sim 30 \tau_{\rm 
turb}$). Therefore, we may conclude that gas evaporation is not the dominant 
destructive physical process for our clumps. In our simulations, though turbulence is 
the main process for the formation of the clumps, turbulent diffusion is also the 
main disruptive mechanism.

Turbulent disruption of clumps may be caused by ablation, fragmentation by strong shocks 
and surface instabilities, such as the Kelvin-Helmoltz instability that rises at the 
surface of the dense core as less dense plasma flows around it (see Pittard, Hartquist \& 
Falle 2010 and references therein). This effect is greatly 
increased in supersonic cases. All these process occur simultaneously in the present 
simulation.

\section{Discussion}

In this work we present MHD numerical simulations of an 
isothermal turbulent ISM gas. We simulate the clumps formed by the turbulent supersonic 
gas and study their dynamical evolution, and the dependence of their 
physical properties with the magnetization of the gas. As a result we observe a large 
dispersion of clump lifetimes and their sizes/masses. However, most importantly, we 
showed that the dense cores survive for timescales larger than the sonic or Alfvenic 
crossing times ($t_{\rm life} \sim 1 - 10 \tau_{\rm s}$ or $\sim 1 - 20 \tau_{\rm A}$). 
The origin of the dense cores 
in the simulation is attributed to shocks. A linear relation between the lifetime of the 
clumps and their column density was found. It may be interpreted as a direct consequence 
of the gas accumulation in a supersonic converging flow.

The most important single result of this study is the linear relation of the clump life time and the column 
density of the clump. This relation shows that the life time of a clump may be 
substantially longer than the sonic or Alfv\'en crossing time of the cloud. One may 
speculate therefore that the 
actual clumps of interstellar gas and even molecular clouds exist for time scales much 
longer than the crossing time. We may remind the reader that the controversy about the life time of molecular 
clouds usually boils down that the magnetic support of the clouds is predicted to 
provide the cloud life time of the order of $10 t_A$, while it is usually assumed that the transient turbulent 
formation of clumps implies the lifetime of the order of minimal of $1t_s$ or $1t_A$. Our 
simulations show that this is not true and that the correlation times of the large scale flows play an important 
role in preserving the clouds. This conclusion is based on isothermal simulations and one 
may be concerned about its relevance in a different scenario. If the flows are not 
isothermal, as in our computation, the density of the clumps created would be different 
since the contrast of density decreases as the cooling becomes less efficient. However, 
the main result in terms of the correlation time of the flows should remain. This because 
still the converging flows that create clumps would be correlated to the turnover time of 
the large scale eddies. Another initial parameter that may possibly cause 
different results is the Mach number of the injected turbulence. In a low Mach number 
case, the origin of the dense regions may be different. Mostly due to the 
density perturbations by the slow and fast waves instead of accumulated 
material within a shock layer (Cho \& Lazarian 2003). In this case, Eq.8 is expected to 
be no longer valid. A similar difference 
may arise when comparing strongly magnetized to the weakly magnetized supersonic turbulence. This will be 
addressed in a future work.

This result may be compared to previous works. Vazquez-Semadeni et al.(2005) studied the 
evolution of cloud cores in subcritical, 
strongly magnetized and supersonic turbulent flows. In their simulations most of 
the dense cores were seen to be extremely transient in nature, with lifetimes ranging in 
the range $0.01 - 0.075 t_{\rm s}$, with exception to a few clumps that became 
gravitationally bound and survived for few local free-fall times. A possible explanation 
for the discrepancies shown in lifetimes is the lower numerical resolution. 
Vazquez-Semadeni et al.(2005) used $256^3$ and $128^3$ cubes in their analysis. Also, the 
shock-capturing method, and the reconstruction method of the variables, may play an 
important role in the diffusion of condensations. In this sense, more numerical tests 
with finer numerical resolutions and better solvers are mandatory. 

Galvan-Madrid et al.(2007) extended this work and studied the lifetimes of cores in 
pre-stellar phases. The prestellar lifetime of a core ($t_{\rm pre}$) is defined as the 
interval between the time when it is first detected at a given threshold and the time 
when its peak density reaches the saturated value indicating the further collapse and 
star formation. In this case, these authors obtained $t_{\rm pre}$ in the range of $\sim 
3 - 10 \tau_{\rm ff}$, which is similar to the ones obtained in the present work.

Interestingly enough, the clumps we observe are very dynamical. Therefore, again extrapolating to molecular clouds,
 we may speculate that the issue of "support of turbulence" in molecular clouds may be 
artificially overemphasized. As long as the turbulent gas is being pushed into molecular 
cloud, it remains turbulent. 
The exception may be presented in the case of small clumps, over which scale the effects 
of viscosity are important. Such clumps can accept a flow of matter from a large scale, at which the viscosity is 
important, but they are small enough to allow development of instabilities within the 
clump.

It is interesting that the life time of clumps scales with the column density. Larson's 
scalings predict that clouds of different sizes tend to present a constant column 
density, though a few authors propose this to be an observational artifact (see 
V\'azquez-Semadeni, Ballesteros-Paredes \& Rodr\'iguez 1997). 
We could interpret this result as the existence of the preferable life time of the 
clumps in a flow. Naturally, in the turbulent fluid such scale is the scale of turbulence
injection at large scale, namely, $L/v_L$. 

The role of magnetic field in star formation is a subject of hot debates. The approaches vary from statements on
 the absolute dominance of magnetic fields (see Shu, Adams \& Lizano 1987) to its 
marginal importance in the presence of turbulence (see Padoan \& Nordlund 1999). 
Our results show that one may expect an increase of magnetic to gaseous pressure as 
turbulence creates clumps. 
Obviously, this result depends on the physical assumptions made in these 
simulations. If the gas was not isothermal, and 
if the cooling process was inneficient, $\beta$ should not increase with the clump 
density as much as we observed here. 
We believe that "reconnection diffusion" based on the ability of magnetic fields to 
reconnect in turbulent media (Lazarian \& Vishniac 1999, Lazarian, Vishniac \& Cho 2004) enabling diffusion 
of magnetic field in relation to conducting fluid, which is the process that can dominate 
ambipolar diffusion in many astrophysical environments (Lazarian 2005, Santos de Lima et al. 2009). 
Rapid magnetic diffusion happens in numerical codes, although the nature of numerical 
diffusion is different from the reconnection diffusion in astrophysical circumstances. Therefore, while numerical 
simulations may probably reproduce features of the astrophysical reality, the degree of
their correspondence is an issue of further studies. An encouraging fact is, however, that as "reconnection diffusion" 
depends on the large scale turbulent eddies, the reproduction of the entire cascade in 
the numerical simulations is not necessary. This work also justifies why one can study 
clumps with one fluid code, i.e. without including ambipolar diffusion effects. 

We show that the degree of elongation of clumps can vary substantially. Identification of 
clumps and 
their evolution, including changes of the clump shapes, requires further studies.

Our present study does not include self-gravity and therefore it cannot be formally applied to 
the self-gravitating Giant Molecular Clouds (GMCs). However, one can get an insight on 
what is going on with GMCs and why they can survive more than the crossing time. Indeed, our results 
are suggestive that the accumulation of matter in GMCs happens on the time scale of the 
large scale turbulent flow, $L/v_L$, where $L$ is injection scale of the turbulent motions and $v_L$ is the 
velocity at the injection scale. Taking $L\approx 100$ pc and $v_L\approx 10$ km/s one 
gets the time scale $\sim 10^7$ years, which is a very crude estimate taking into account 
the approximate nature of the estimates\footnote{Observational studies of turbulence using VCA and VCS techniques 
(see Lazarian 2009 and ref. therein) should provide more precise estimates of the 
turbulent velocities.}, though it is in agreement with the observational estimative of 
Blitz \& Shu (1980). On 
this timescale the GMCs are likely be dispersed through the feedback from star formation 
and star evolution, which resembles a pre-turbulence paradigm of GMC evolution. In 
addition, there is no problem of explaining the existence of turbulence in GMCs, as in this picture they are being
 formed all the time of their existence rather than being static entities which are 
relaxing their internal motions. While the turbulent formation of GMCs is not a new concept 
(see Ballesteros-Paredes et al. 1999) we believe that our study of clump evolution 
provides it with more solid ground and justification.

\section{Summary}

In the paper above we obtained the following results:

1. The magnetization of clumps, compared to the thermal pressure, increases with the mass 
of the clumps. 

2. The lifetime of a clump scales is not limited by the sound crossing time on the scale of the clump, but is determined by the existence of the large-scale converging flow with clumps surviving 
many Alfv\'en and sonic crossing times.

3. Our model of clumps predicts that the lifetime of clumps scales linearly with the column density, which
corresponds well to our numerical simulations.

\acknowledgments

The authors thank the anonymous referee for helping in the improvement of this paper. 
A.L. and D.F.G. thank the financial support of the Center for 
Magnetic Self-Organization in Astrophysical and Laboratory Plasmas, NSF grant AST 0808118 and the Brazilian 
agency FAPESP (No.\ 2009/10102-0).

\end{document}